\documentclass[a4paper]{jpconf}
\usepackage{graphicx}
\begin{document}
\title{CPT and Lorentz violation as signatures for Planck-scale physics}

\author{Ralf Lehnert}

\address{Instituto de Ciencias Nucleares,
Universidad Nacional Aut\'onoma de M\'exico,
A.~Postal 70-543, 04510 M\'exico D.F., Mexico}

\ead{ralf.lehnert@nucleares.unam.mx}

\begin{abstract}
In recent years, 
the breakdown of spacetime symmetries 
has been identified 
as a promising research field 
in the context of Planck-scale phenomenology. 
For example, 
various theoretical approaches   
to the quantum-gravity problem 
are known to accommodate 
minute violations of CPT invariance.
This talk 
covers various topics within this research area. 
In particular, 
some mechanisms for spacetime-symmetry breaking 
as well as the Standard-Model Extension (SME) 
test framework 
will be reviewed; 
the connection between 
CPT and Lorentz invariance 
in quantum field theory 
will be exposed;
and various experimental CPT tests 
with emphasis on matter--antimatter comparisons 
will be discussed.
\end{abstract}

\section{Introduction}

The discrete spacetime transformations of 
charge conjugation (C),
parity inversion (P), 
and time reversal (T), 
as well as various combinations like CP and CPT 
have played a key role in fundamental physics 
since the 20th century. 
For example, 
prior to the 1950s, 
parity was widely believed to be an exact symmetry of nature.\footnote{ 
However, it had been 
suggested several times 
and in different contexts 
that parity might, 
in fact, 
be violated.}
This believe changed when 
convincing experimental evidence 
of parity violation in the beta decay of ${}^{60}$Co was found~\cite{Wu57}.
This discovery paved the way 
for a viable theoretical description of the weak force.
At present, 
it is known that not only P, 
but also other discrete spacetime transformations 
(e.g., T and CP) 
are not associated with exact symmetries in nature. 
Experimental and theoretical studies in this field
have remained an active and important research area, 
which may give insight into physics beyond the Standard Model.

The role of the CPT transformation 
is special in this context. 
While there are no particular theoretical reasons 
why C, P, T, CP, CT, and PT 
should be conserved, 
CPT invariance must hold 
in conventional physics~\cite{cpt}:
the celebrated CPT theorem states 
that under mild assumptions 
Lorentz symmetry implies CPT invariance
in a unitary quantum field theory. 
In other words,
relativity and quantum mechanics 
essentially require CPT to be conserved. 
It follows 
that CPT tests are a tool 
for probing the foundations of physics.
This fact 
together with the ultrahigh sensitivities 
attainable in CPT-violation searches 
have led to a recent revival of interest 
in this subject.

The CPT transformation provides a connection 
between a particle and its antiparticle. 
This suggests 
that CPT conservation 
is associated with a symmetry between matter and antimatter. 
One can prove
that this is indeed the case: 
the magnitudes of 
the mass, charge, decay rate, gyromagnetic ratio, 
and other intrinsic properties of a particle 
are identical to those of its antiparticle 
if CPT invariance holds. 
Such arguments can also be applied to 
systems of particles and their dynamics. 
For example, atoms and their corresponding anti-atoms 
must have the same spectra, 
and a particle-reaction process and its CPT conjugate 
must exhibit equal reaction cross sections. 
It follows that experimental matter--antimatter comparisons 
can serve as probes for the validity of CPT symmetry.

However, 
CPT symmetry can also be tested 
in other systems.
The basic idea is 
that if CPT is violated 
one of the assumptions 
necessary to prove the CPT theorem must be relaxed. 
In the context of axiomatic field theory, 
one can prove rigorously 
that CPT violation implies Lorentz breakdown 
if quantum-mechanical probability conservation 
is to be maintained~\cite{anticpt}.\footnote{For another approach 
to CPT breaking resulting from apparent probability nonconservation,
see N.~Mavromatos' and S.~Sarkar's contributions to these proceedings.}
We remark, 
however, 
that the converse of this fact---namely that Lorentz violation implies 
CPT breaking---is not true in general. 
In any case, it follows 
that in conventional quantum mechanics,
CPT-violation searches are at the same time Lorentz-symmetry tests.
This offers the possibility of probing CPT invariance 
via certain experiments 
designed to test Lorentz symmetry.

The present talk develops these particular ideas further. 
We begin by reviewing the theoretical motivations 
for considering CPT violation 
and the associated Lorentz breakdown. 
We then recall the Standard-Model Extension (SME)---the framework 
for the description 
of CPT- and Lorentz-violating effects 
at low, presently attainable energies. 
The final part discusses a few possibilities for testing these ideas.

\section{Motivations and model for CPT and Lorentz violation}

For the identification and analysis 
of experimental CPT- and Lorentz-violation searches, 
a test framework is needed. 
This framework must allow for small departures 
from exact CPT and Lorentz symmetry.
The SME mentioned above 
is such a framework.
It has been developed over the last decade, 
and it is constructed 
to be relatively general and independent 
of the (thus far unknown) details of the underlying physics~\cite{sme}. 
At the same time,
the SME maintains numerous desirable features 
of conventional physics.
We begin this section 
by sketching the line of reasoning 
that has been employed to establish the SME.

In a first step, 
it is necessary to decide how to implement 
CPT and Lorentz violation into a test model. 
Features associated with 
the requirement of coordinate independence 
provide one possible basis 
for classifying departures from Lorentz symmetry.
Coordinates, 
which are a pure product of human thought, 
label spacetime points in a largely arbitrary way; 
they are descriptive tools 
and as such they lack physical reality. 
Model predictions 
must therefore remain independent 
of the chosen coordinates. 
This can be achieved by working on a spacetime
manifold and representing physical quantities by geometric objects like
tensors or spinors~\cite{coordindep}.
However, this principle does not fix the type
of the underlying manifold: 
for example, 
both Lorentzian and Galilean spaces
would equally be consistent with coordinate independence. 

The above reasoning 
suggests one possibility 
for the implementation of Lorentz breakdown:
the underlying spacetime structure 
is no longer Lorentzian, 
so that inertial frames are not connected 
via the usual Lorentz transformations.
In other words, 
covariance under Lorentz transformations
is replaced by 
covariance under some other symmetry transformation.
Such deformations of Lorentz symmetry 
have been discussed in the literature. 
However, 
their interpretation---and in particular their CPT properties---remain 
unclear at present. 
These ideas are not directly employed 
in the construction of the SME.

A second possibility for the implementation of CPT and Lorentz violation 
maintains the conventional Lorentzian spacetime structure 
and employs a nontrivial vacuum instead. 
Such a vacuum can be described by 
a nondynamical tensorial background; 
such a background could lead 
to direction- and boost-dependent physics, 
for example.
This situation is somewhat similar 
to electrodynamics in macroscopic media: 
covariant behavior under Lorentz transformations
of the Minkowski frame (i.e., coordinate independence)
is maintained, 
but the propagation of light need not be isotropic 
and can be slower that $c$.
The SME incorporates this type of CPT and Lorentz violation.

Such a nontrivial vacuum with CPT and Lorentz violation 
can now be implemented in a model Lagrangian, 
which is to be interpreted as a low-energy effective field theory.
The motivation for this approach is the following. 
Effective field theories have been exceptionally successful 
in particle, nuclear, and condensed-matter physics. 
In fact, 
the conventional Standard Model (SM) and General Relativity (GR)
are usually viewed as effective field theories;
leading-order CPT- and Lorentz-breaking corrections 
{\em outside} the framework of effective field theory 
therefore seem somewhat contrived. 
Moreover, 
this approach can naturally incorporate 
practically the entire basis of known physics---the SM and GR---as limiting cases. 
These two features 
(i.e., a theoretically well understood framework 
containing all of present-day physics) 
ensure broadest applicability of the SME 
for all currently feasible experiments.

With the above considerations, 
the SME Lagrangian ${\cal L}_{\rm SME}$ now takes the form
\begin{equation} 
\label{SMElagr}
{\cal L}_{\rm SME}={\cal L}_{\rm SM}+{\cal L}_{\rm GR}+\delta{\cal L}_{\rm SME}\;.
\end{equation}
Here, 
${\cal L}_{\rm SM}$ and ${\cal L}_{\rm GR}$ denote 
the Lagrangians of the SM and GR, 
respectively. 
It is thus evident 
that the SME incorporates 
the entire foundation of established physics,
as argued above.
The CPT- and Lorentz-violating corrections 
are contained in $\delta{\cal L}_{\rm SME}$, 
which is formed by 
contracting the stipulated vacuum background tensors
with SM or GR fields. 
For example, 
$\delta{\cal L}_{\rm SME}$ includes the term 
$b_\mu\overline{\psi}\gamma^5\gamma^\mu\psi$, 
where $\psi$ is, e.g., the electron field of the SM. 
The background vector $b_\mu$
is assumed to be caused by underlying physics.
This sample term violates both CPT and Lorentz symmetry, 
and it is $b_\mu$ that experiments can measure or constrain.
We remark that 
physically desirable features, 
such as power-counting renormalizability and 
SU(3)$\times$SU(2)$\times$U(1) gauge invariance,
are often imposed in the literature;
this special case is then referred to as the minimal SME. 
Various studies 
have solidified the theoretical foundations
of the minimal SME~\cite{smetheory}.

Thus far, 
the SME has been constructed by hand 
without reference to physics beyond the SM. 
In the remaining part of this section, 
we list a few mechanisms in underlying physics 
that can generate the tensorial backgrounds
contained in the SME, 
which are responsible for violating CPT and Lorentz invariance.
 
{\em Spontaneous CPT and Lorentz breakdown in string theory.}---From 
a theoretical viewpoint, 
spontaneous symmetry violation (SSV) provides 
an attractive mechanism for 
generating CPT and Lorentz breakdown. 
SSV is well established in solid-state physics, 
and in the electroweak model 
it is believed to be responsible for creating the masses of elementary particles. 
The essence of SSV is that 
a symmetric zero field value  
does not correspond to the lowest-energy state. 
Instead, 
non-vanishing vacuum expectation values (VEVs) 
are energetically preferred. 
In string field theory, 
one can demonstrate that 
SSV can produce VEVs of vector and tensor fields, 
which can then be related to the CPT- and Lorentz-violating 
background in the SME 
(e.g., $b_\mu$ in the example in the previous paragraph)~\cite{ssv}.

{\em Nontrivial spacetime topology.}---The basic idea 
behind this approach 
is the possibility that 
one of the usual three spatial dimensions 
is compactified. 
On observational grounds, 
the compactification radius $R$ would clearly have to be very large. 
In any case, 
the local structure of flat Minkowski space is preserved. 
The finite size of the compactified dimension implies periodic boundary conditions, 
which in turn lead to a discrete momentum spectrum, 
so that a Casimir-type vacuum emerges. 
It is then intuitively reasonable that 
a vacuum of this type 
can possess a preferred direction 
along the compactified dimension. 
Indeed, 
one can show 
that such a situation is described effectively by
certain SME terms~\cite{topology}; 
the corresponding background vectors
are inversely proportional 
to the compactification radius $R$.

{\em Cosmologically varying scalars.}---A varying scalar, 
such as a varying coupling or a cosmological scalar field, 
is typically associated with the breakdown of translational symmetry.
This feature occurs regardless of the mechanism 
responsible for the spacetime dependence.
Since translations are closely intertwined with 
the Lorentz transformations 
within the Poincar\'e group, 
it is unsurprising that 
the translation-symmetry violation 
can also affect Lorentz invariance. 
Consider, for example, a physical system 
with a varying coupling $\alpha(x)$ 
and two scalar fields $\phi$ and $\Phi$.
Suppose further that
the Lagrangian for this system 
contains a kinetic-type term
of the form $\alpha(x)\,\partial^\mu\phi\,\partial_\mu\Phi$.
A suitable integration by parts then generates the term 
$-(\partial^\mu\alpha)\,\phi\,\partial_\mu\Phi$
while leaving unaffected the equations of motion, 
and thus the physics. 
If the variation of $\alpha(x)$
is slow (say on cosmological scales),
the gradient $(\partial^\mu\alpha)$ 
can be taken approximately constant 
on laboratory scales. 
It is then apparent 
that the external nondynamical gradient
can be identified with one of the background vectors 
in the SME~\cite{scalars}. 

{\em Non-commutative field theory.}---This popular 
approach to physics beyond the SM 
postulates that 
coordinates are no longer real numbers, 
but rather operators 
satisfying nontrivial commutation relations 
such as $[x^\mu,x^\nu]=i\theta^{\mu\nu}$. 
The quantity $\theta^{\mu\nu}\neq0$ 
is often taken as spacetime constant 
and selects preferred directions in the non-commutative space. 
To interpret such models physically, 
one can transform them into ordinary field theories 
via the Seiberg--Witten map. 
The resulting field theory on conventional Minkowski space 
still contains the nondynamical constant $\theta^{\mu\nu}$, 
which acts as a background tensor,
i.e., 
SME terms are generated~\cite{noncomm}.

{\em Loop quantum gravity.}---Another widely known approach
to a quantum version of General Relativity 
is loop quantum gravity. 
In semiclassical calculations, 
various results have been derived 
that indicate Lorentz violation 
in electrodynamics 
and for fermions
under certain reasonable assumptions~\cite{LQG}. 
An effective description of these effects 
is contained in the SME.

\section{Experimental CPT-violation searches}

The SME framework discussed in the previous section 
can now be employed to predict experimental signatures 
for CPT and Lorentz violation. 
For example, 
a key concept in the context of Relativity theory
is the speed of light and its constancy;
the SME predicts deviations from this concept 
that can be searched for experimentally~\cite{ktr}.
But CPT and Lorentz symmetry 
also provide the basis for many other properties 
in numerous physical systems. 
Accordingly, 
a large number of additional CPT and Lorentz tests 
have been analyzed within the context of the SME~\cite{tables}.
In this section, 
we review a few of these ideas 
with focus on those tests 
that not only bound Lorentz breaking
but also CPT violation.

{\em Spectropolarimetry of cosmological sources.}---The photon sector 
of the minimal SME 
contains one type of coefficient 
that violates both CPT and Lorentz symmetry, 
the Chern--Simons-type $(k_{AF})^\mu$ term. 
For example, 
this term leads to birefringence in the propagation 
of electromagnetic waves~\cite{kAF}, 
vacuum Cherenkov radiation~\cite{cherenkov}, 
and shifts in cavity frequencies~\cite{cavity}. 
These are effects
that can be searched for experimentally.
Birefringence studies in cosmic radiation are particularly well suited 
because the extremely long propagation distances 
translate into ultrahigh sensitivity 
to this type of Lorentz and CPT violation.
An analysis of experimental data 
from cosmological sources 
has yielded a limit on $(k_{AF})^\mu$ 
at the level of $10^{-42}$GeV~\cite{kAF}.

{\em Studies involving cold antihydrogen.}---Comparisons 
of the spectra of hydrogen (H) and antihydrogen ($\overline{\rm H}$) 
are well suited for CPT and Lorentz tests. 
Among the various transitions 
that can be considered, 
the unmixed 1S--2S transition appears to be an excellent candidate: 
its projected experimental resolution is expected to be about 
one part in $10^{18}$, 
which is promising in light of potential Planck-suppressed quantum-gravity effects. 
On the other hand, 
the corresponding leading-order SME calculation 
establishes identical shifts for free H or $\overline{\rm H}$ 
in the initial and final states 
with respect to the conventional energy levels. 
From this perspective, 
the 1S--2S transition is actually less suitable 
for the measurement of unsuppressed CPT- and Lorentz-violating signals.
The largest non-trivial contribution to this transition 
within the SME test framework 
is produced by relativistic corrections, 
and it is multiplied by two additional powers of the fine-structure parameter $\alpha$. 
The expected energy shift, 
already at zeroth order in $\alpha$ expected to be minuscule, 
is therefore associated 
with an additional suppression factor 
of more than ten thousand~\cite{antiH}. 

Another transition 
that can be employed for CPT- and Lorentz-violation searches 
is the spin-mixed 1S--2S transition.
When H or $\overline{\rm H}$ is confined with magnetic fields---such as in a Ioffe--Pritchard trap---the 1S and the 2S levels are each split 
due to the usual the Zeeman effect. 
In the framework of the SME, 
one can demonstrate 
that in this case the 1S--2S transition 
between the spin-mixed states is indeed shifted 
by CPT and Lorentz breaking 
at leading order. 
A disadvantage from a practical viewpoint 
is the $\vec{B}$-field dependence of this transition, 
so that the experimental sensitivity is limited 
by the size of the inhomogeneity 
in the trapping  magnetic field. 
The development of novel experimental techniques 
might circumvent this issue, 
and resolutions close to the natural linewidth
might then be achievable~\cite{antiH}. 

A third transition suitable for CPT- and Lorentz-violation searches
is the hyperfine Zeeman transitions within the 1S state.\footnote{See also 
E.~Widmann's contribution to these proceedings.} 
Even in the limit of a vanishing magnetic field, 
the SME predicts leading-order effects for two of the transitions
between the Zeeman-split states.
We mention 
that this result may also be practical 
from an experimental point of view
because various other transitions of this type 
(e.g., the conventional Hydrogen-maser line)
can be well resolved in measurements~\cite{antiH,hyperfine}. 

{\em Tests in Penning traps.}---The SME predicts not only 
that atomic energy levels can be affected 
by the presence of CPT and Lorentz violation, 
but also, e.g., 
proton and antiproton levels in Penning traps. 
A perturbative calculation establishes 
that only one SME coefficient 
(a $b^\mu$-type background vector
mentioned in the previous section)
contributes to the transition-frequency difference
between the proton and its antiparticle
at leading order.
More specifically, 
the anomaly frequencies are shifted in opposite directions 
for protons and antiprotons. 
This effect can be employed to 
extract a clean experimental bound 
on the proton's $b^\mu$~\cite{penning}.

{\em Neutral-meson interferometry.} 
A widely known standard CPT test 
involves the comparison of the K-meson's mass 
to that of its antimeson:
even very small mass differences 
would be measurable in Kaon-oscillation experiments.
Although the SME contains only one mass parameter 
for a given quark species and the corresponding antiquark species, 
these particles are nevertheless affected differently 
by the CPT- and Lorentz-violating background in the SME. 
This allows the dispersion relations for mesons and antimesons to differ, 
so that mesons and antimesons can have distinct energies. 
It is this difference in energy 
that ultimately affects interferometric experiments
and is therefore potentially observable in such systems~\cite{mesons,kloe}. 
Note that not only the K-meson 
but also other neutral mesons can be studied.
Note also that besides CPT violation, 
Lorentz breaking is involved as well, 
so that boost- and rotation-dependent signals 
can be searched for.\footnote{See also 
A.~Di Domenico's contribution to these proceedings.} 

\section*{Acknowledgments} 
The author wishes to thank the organizers 
for arranging this stimulating meeting 
and for subsidizing my attendance.
This work was also supported in part by 
the European Commission 
under Grant No.\ MOIF-CT-2005-008687
and by CONACyT under Grant No.\ 55310.

\end{document}